\documentclass[showkeys]{revtex4}
\usepackage{textcomp}
\usepackage{multirow}
\usepackage{amsmath}
\usepackage{amssymb}
\usepackage{graphicx}
\usepackage{epstopdf}
\usepackage{float}
\usepackage{booktabs}
\usepackage{dcolumn}
\usepackage{bm}

\begin{document}

\title{Expectation on probing the origin of the cosmic ray knee with the LHAASO experiment}

\author{Chao Jin}
\email{jinchao@ihep.ac.cn}
\affiliation{Key Laboratory of Particle Astrophysics, Institute of High Energy Physics, Chinese Academy of Sciences,Beijing 100049,China}
\affiliation{University of Chinese Academy of Sciences, 19 A Yuquan Rd, Shijingshan District, Beijing 100049, P.R.China}
\author{Li-Qiao Yin}
\affiliation{Key Laboratory of Particle Astrophysics, Institute of High Energy Physics, Chinese Academy of Sciences,Beijing 100049,China}
\affiliation{University of Chinese Academy of Sciences, 19 A Yuquan Rd, Shijingshan District, Beijing 100049, P.R.China}
\author{Song-zhan Chen}
\affiliation{University of Chinese Academy of Sciences, 19 A Yuquan Rd, Shijingshan District, Beijing 100049, P.R.China}
\author{Hui-Hai He}
\affiliation{University of Chinese Academy of Sciences, 19 A Yuquan Rd, Shijingshan District, Beijing 100049, P.R.China}

\begin{abstract}
The cosmic-ray (CR) knee and the compositions contain abundant information for probing the CR's origin, acceleration and propagation mechanisms, as well as the frontier of the fundamental physics. Realizing that major proposals toward the knee's shape can be divided into two categories: the rigidity-dependent (also Z-dependent) knee and the mass-dependent (also A-dependent) knee, where the former one relates to the acceleration or the propagation mechanisms, and the other one is often associated with the new physics, it is essential to precisely measure the individual compositions. Benefit from the high altitude and hybrid detection methods, the LHAASO experiment has the ability in determining the individual component and brings us an opportunity in discriminating these two models. We test this expected ability of LHAASO from 100 TeV to 10 PeV with 3-year observation. And find the dominant component at the knee is essential to this issue, while much heavier nuclei occupying the knee leads to higher significance. In the analysis, the He-dominant knee under the A-dependent case can be recognized at the significance about 6.6 $\sigma$, while the P-dominant knee under the Z-dependent case will be classified with 2 $\sigma$ significance.
\end{abstract}
\maketitle

\section{Introduction}
The CRs' spectrum, and their compositions, contains abundant information for probing their origin, acceleration and propagation mechanisms. The spectral break of the CR called "knee" was first discovered in 1959 \cite{1959JETP...35....8K}, but its origin still remains a puzzle over nearly 60 years. Although current measurements about the knee are different among the ground-based experiments under different conditions, they can come into an agreement assuming the 15\% uncertainty of the energy reconstruction \cite{2003JPhG...29..809H}. And it is summarized that the knee is described by a resembled double-power law, with the spectral index -2.7 below 4 PeV, and steepening to -3.1 above this energy \cite{2013AIPC.1516..185H}.
\par
In order to explore the origin of the knee, precise measurement about the spectra of individual compositions is important. Due to the low-flux level of the CR knee about only 1 $m^{-2} yr^{-1}$, current measurements are settled ground-based to obtain the larger detecting aperture. And they are all performed through the indirect method, i.e. the Monte Carlo simulation about the extensive air shower (EAS) induced by the CRs impinging on the atmosphere. In lack of the knee energy data from the accelerator, the adopted hadronic interaction model extrapolated from the lower energy may bring unknown systematic errors in determining the spectrum and the compositions. There have been much efforts spent on the component measurement, but results make large discrepancy. Measurement from KASCADE favours that the knee of the all-particle spectrum is contributed by the steepening of the light species \cite{2005APh....24....1A}, and a heavy knee (mainly by the iron) is observed around 80 PeV \cite{2011PhRvL.107q1104A}. But the ARGO+WFCT finds the spectral break of the light species appears at $\sim$ 700 TeV \cite{2015PhRvD..92i2005B}. Meanwhile, in the analysis with various experiments, it is found the dominant specie at the knee is likely to be the Helium \cite{2006JPhG...32....1E}, and measurement from GAMMA attribute the knee to the light species (P and He) \cite{2014PhRvD..89l3003T}. Different from those, the $AS\gamma$ experiment indicates the knee originates from the nuclei higher than the Helium \cite{2006PhLB..632...58T}.
\par
With the poor information about the individual components, various explanations about the the knee's origin have been proposed (see \cite{2004APh....21..241H} and references therein). From the point of view about the astrophysical origin, many proposals attribute it to the change of the acceleration mechanism \cite{2007ApJ...661L.175B, 2002PhRvD..66h3004K, 2004A&A...417..807V}, the single source contribution \cite{2009arXiv0906.3949E, 2014PhRvD..89l3003T}, or the propagation effect in the Galaxy \cite{1995ICRC....2..697S, 1993A&A...268..726P, 2001ICRC....5.1889L, 2014PhRvD..90d1302G, 2015PhRvD..91h3009G}. In the consideration of the interaction effect, diverse models are proposed including the new channel of the hadronic interaction model \cite{2010EPJC...68..573D, 2001ICRC....5.1760K} and collision with exotic particles \cite{2003GReGr..35.1117K, 2008JCAP...12..003M, 2009JCAP...06..027B, 2003APh....19..379W, 2016arXiv161108384J} (relic neutrinos, the Dark Matter, SUSY, graviton, etc.). Besides, the regular process such as pair-production and disintegration of photons \cite{2009ApJ...700L.170H, 2001ICRC....5.1979T, 2002APh....17...23C} also belong to such kind of model. By investigating these proposals, it is found that most of them can be divided into two categories, the Z-dependent knee and the A-dependent knee (A, Z correspond to the CR nuclei's atomic number and charge respectively). The former one mainly relates to the CR's acceleration and propagation mechanisms, while the other one indicates many of them associate with the new physics processes. Thus, distinguishing between these two model is essential for exploring both the fundamental problems of CR and the new physics.
\par
The LHAASO experiment \cite{LHAASO_review} is the next generation of the ground-based experiment located at high altitude of 4410 m, at which the EAS induced by nuclei around the knee develops to the maximum and is expected to have less dependence on the hadronic interaction model. On the other hand, the LHAASO experiment combines the hybrid detection method, including detecting the charged particles, muons, as well as the Chrenkov/fluorescence photons. The charged particles construct the major part of the EAS' lateral distribution which is useful in determining the arrival directions, core locations, and primary energies, while the collected Cherenkov image is a good estimator about the CR energy and also sensitive to the CR component. The muons, generated by the decay of the charged pions, depend on the primary mass of CRs and have the ability in recognition of the primary CR species as well.
\par
Benefit from those advantages, the forthcoming LHAASO experiment will bring us an opportunity on the precise measurement about individual CR compositions. In this work, we investigate the capability of LHAASO in distinguishing the Z-dependent and A-dependent knee models. The contents are organized as follows: the section II contains the brief information about LHAASO, the section III contains the detail procedure of the analysis and the results of both the Z-dependent and the A-dependent knee models. In the last section, a conclusion and discussion is delivered.

\section{The LHAASO experiment}
\begin{center}
\begin{figure*}
\centering
\includegraphics[width=.45\textwidth]{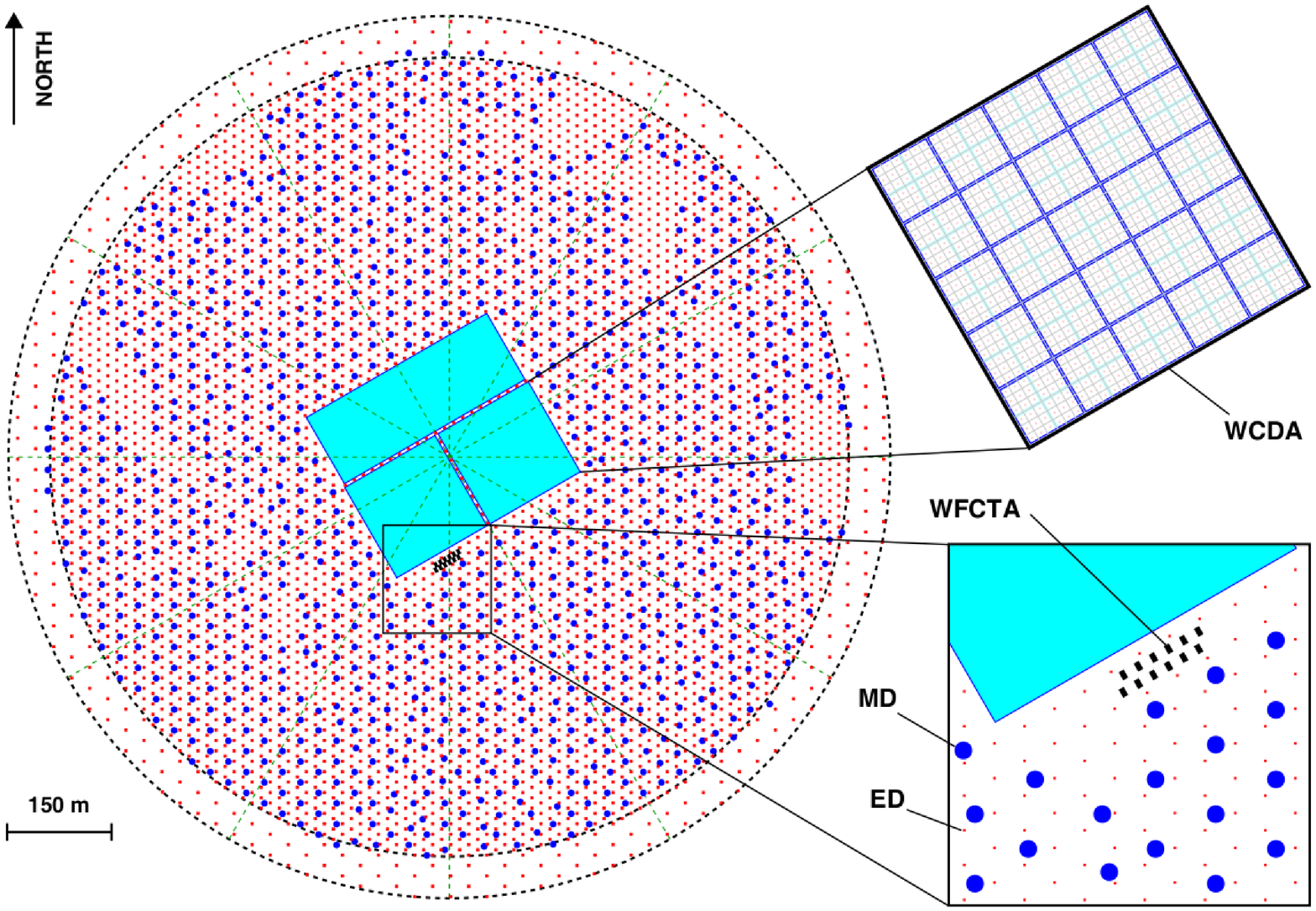}
\caption{Layout of the LHAASO experiment. The insets show the details of one pond of the WCDA, and the EDs (red points) and MDs (blue points) of the KM2A. the WFCTA, located at the edge of the WCDA, is also shown}
\label{fig:lhaaso}
\end{figure*}
\end{center}

%The LHAASO is located at Sichuan.
%\par
%The major detection array consists of three parts, including the KM2A, WCDA, WFCTA.
%\par
%The composition measurement is through the hybrid detection.
The LHAASO experiment will be located at high altitude (4410 m a.s.l.) in the Daocheng site, Sichuan Province, China. It consists of an EAS array KM2A covering 1.3 $km^2$ area, 78000 $m^2$ closed packed water Cherenkov detector array (WCDA), and 12 wide-field Cherenkov/fluorescence telescopes (WFCTA). The KM2A is composed by two sub-arrays, including 1 $km^2$ array of 5195 electromagnetic particle detectors (ED) and the overlapping 1 $km^2$ array of 1171 underground water Cherenkov ranks for muon detectors (MD). The WCDA contains three water ponds with the effective depth about 4 m. Each pond is divided by 5 m $\times$ 5 m cells with an 8-inch PMT located at the bottom center to watch the Cherenkov light generated by the EAS secondary particles in the water. And the focal plane camera in each telescope of WFCTA has 32 $\times$ 32 pixels with every pixel size $0.5 \ \times \ 0.5$. The layout of each component of LHAASO is illustrated in the Fig. \ref{fig:lhaaso}.
\par
The LHAASO experiment aims at measuring precise primary CR spectrum through the hybrid detection method from 10 TeV to EeV, together with a sensitivity of 1\% Crab unity in order to survey the northern hemisphere to identify the gamma ray sources with full duty cycle, and measures the spectra of all the sources simultaneously within a wide energy range between $10^{11}$ and $10^{15}$ eV. The WFCTA will effectively measure the primary energy of different mass groups and will be operated in three modes covering wide energy range. The first and second are the Cherenkov mode focusing on the low energy (30 TeV - 10 PeV) and middle energy (10 PeV - 100 PeV) respectively, and the third mode is the fluorescence light detecting with the rearranged array allowing the energy spectrum and composition measurement above 100 PeV.
\par
The CR composition measurement is a challenging task, and there are four parameters from the hybrid detectors above-mentioned crucial in identifying the primary components. One parameter is offered by the WCDA calculating the energy flow near the core, another parameter is offered by the MD counting the total muons, the other two are extracted from WFCTA on behalf of the morphology of the Cherenkov image and the shower maximum $X_{max}$. These parameters utilize both the lateral and longitudinal features of the shower from the EAS secondary particles and Cherenkov photons, and are sensitive to the primary mass group. The detail description of the hybrid detecting method is shown in \cite{lhaaso_yin}. In this work, we focus on the first mode of WFCTA in combination with other arrays in the energy range 100 TeV $\sim$ 10 PeV. In order to insure the purity of the primary component, we test the expected observation about the proton (P) and the light species (P + He), which is sufficient to test the ability of LHAASO to distinguish whether the knee is A-dependent or Z-dependent. And it has been justified that the purity can reach up to 90 \% for P and 95 \% for P + He \cite{lhaaso_yin}.

\section{Statistical Test Of The CR Knee Models}
\subsection{Statistics Analysis Algorithm}
In the analysis, the hypothesis test is implemented to investigate the ability in distinguishing between the CR knee models. There are two hypotheses need to be tested, including the Z-dependent model versus the A-dependent model under the hypothesis of the A-dependent knee, and the A-dependent model versus the Z-dependent model under the hypothesis of the Z-dependent knee. And the chi-square test-statistics $\triangle \chi^2$ \cite{2018PhLB..780..181F} is formulated as
\begin{equation}
\label{test_statistics}
\triangle \chi^2 = min \chi^2 (H_0) \ - \ min \chi^2 (H_1)
\end{equation}
\par
As the confusing issue is whether the CR knee is Z-dependent or A-dependent, the $H_0$ and $H_1$ in Eq. \ref{test_statistics} represent these two hypotheses. The $H_1$ is the real explanation about the knee, and $H_0$ is the imaginary one. Noted that a better fitting result tends to the lower-value of the chi-square, the test-statistics $\triangle \chi^2$ will have a positive mean value only if the $H_1$ hypothesis is preferred. The chi-square of each model H is constructed by combining both the P and light components (P + He) measurements through
\begin{equation}
\label{chi_square}
\chi^2 (H) = \sum_i \left( \frac{(\mu_i(P|H) - N_i(P))^2}{\sigma_{stat}^2(N_i(P))} + \frac{(\mu_i(P+He|H) - N_i(P+He))^2}{\sigma_{stat}^2(N_i(P+He))} \right)
\end{equation}
Where $N_i(P)$ and $N_i(P+He)$ are the observed event counts of each bin, and $\mu_i(P|H)$ and $\mu_i(P+He|H)$ are fitted corresponding to the expectations of each model. Currently only the statistical error is considered, while the influence from the systematic uncertainty will be discussed in the last section. To minimise the chi-square with respect to each model, the MINUIT algorithm from the Root package \cite{Root_Minuit} is implemented.
\par
In the evaluation for the expectations of each bin, detection responses are adopted through
\begin{equation}
\label{bin_counts}
\mu_i = \Phi_i \cdot \triangle E_i \cdot T \cdot Aper_i
\end{equation}
where $\Phi_i$ is flux from the model prediction, $Aper_i$ is the aperture of the hybrid detection from the LHAASO and is adopted from \cite{lhaaso_yin}. T is the total exposure time and it is set to 3 years with 10\% duty cycle of the WFCTA. As mentioned above, the hybrid detection from LHAASO results in a low contamination rate, which is only around 5\% about the light components and 10\% about the proton. Assuming the contamination can be subtracted appropriately, it is expected to have little influence and can be neglected in the analysis.
\par
Assuming the event count $N_i$ falling in each bin obeys the Gaussian distribution, it is sampled using the Monte Carlo method, whose expectation shape is determined by the empirical function or explicit interaction calculation. In order to determine the distribution of the test-statistics $\triangle \chi^2$, 10000 pseudo experiments under the $H_0$ hypothesis are sampled according to directly fitting to the observation with the $H_0$ model, and the same algorithm in Eq. \ref{test_statistics} is performed for each pseudo experiment. We derive the exclusion significance of the $H_0$ hypothesis by comparing the observation $\triangle \chi^2_{obs}$ with the distribution of $\triangle \chi^2$. Totally 500 observation is generated to obtain an averaged exclusion significance as the evaluation of the LHAASO's capability in discriminating different knee models.

\subsection{A-Dependent Model Analysis}
The mechanism resulting in the A-dependent knee often relates to the threshold interaction. Such an process occurs when the nuclei $m_A$ impinges on the target, and the energy of each nucleon reaches the same threshold. As a result, the spectral break corresponds the threshold energy, and increase with the atomic number A. Our previous work \cite{2016arXiv161108384J} found the similar Lorentz factor $\sim 10^6$ for both the CR knee and electron's TeV cutoff as reported by latest measurement \cite{2015ICRC...34..411S, 2011ICRC....6...47B, 2009A&A...508..561A}, and tried to explore the latent mechanism behind this similarity. An unknown light particle is introduced which is abundant in the Galaxy with the mass less than 1 eV. The CRs will suffer the energy loss when the momentum transfer is sufficient to convert the particle to the heavier one, i.e. $10^6$ times heavier. As a result the thresholds of the CR proton and Helium derived are about 1 and 4 PeV respectively. Considering that the threshold interpretations of the knee often relate to the similar mechanism, we adopt our model as representative of them and denote it as the X model.
\par
The threshold interaction can generate the sharp knee structure by tuning the interaction cross section across the knee. And we depict the sharp knee shape with the double power-law function, where
\begin{equation}
\label{power_law}
\Phi(E) = \begin{cases}
\Phi_0 \left( \frac{E_{cut}}{1 \ TeV} \right)^{\gamma_1} \left( \frac{E}{E_{cut}} \right)^{\gamma_1}, \ E < E_{cut} \\
 \Phi_0 \left( \frac{E_{cut}}{1 \ TeV} \right)^{\gamma_1} \left( \frac{E}{E_{cut}} \right)^{\gamma_2}, \ E > E_{cut}
\end{cases}
\end{equation}
The parameter $\Phi_0$ is the normalized flux. And the knee of each component $E_{cut}$ is $Z \cdot E_0$ under the Z-dependent model ($H_0$) and $A \cdot E_0$ under the A-dependent model ($H_1$).
\par
\begin{center}
\begin{figure*}
\centering
\includegraphics[width=.45\textwidth]{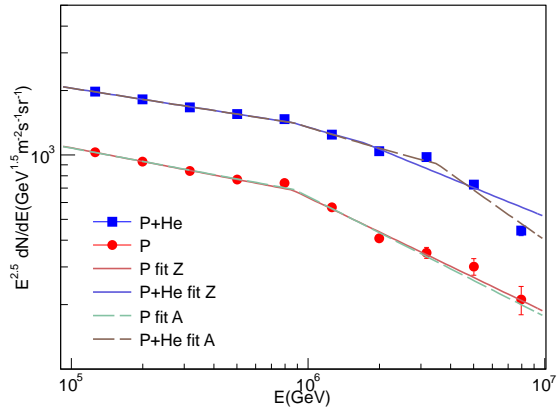}
\caption{One sampled observation and the corresponding fitting results under the X model. The blue dots are the sampled light component spectrum, and the red dots are the sampled proton spectrum. Solid lines are the Z-dependent model ($H_0$) fitting results, and dashed lines are the A-dependent model ($H_1$) fitting result.}
\label{fig:X_model_obs}
\end{figure*}
\end{center}

One sampled observation together with the fitting results of different hypotheses are illustrated in the Fig. \ref{fig:X_model_obs}. Both of model fittings have no difference about the proton spectrum, while the divergence is mainly produced by the light component spectrum at the last three bins by eye, whose energy is above the knee of the He. As the model predicts the knee of He is located at about 4 PeV. The Z-dependent fitting generates the He's knee at about 2 PeV, which is insufficient to explain the higher-energy spectral break of the light components.
\par
\begin{center}
\begin{figure*}
\centering
\includegraphics[width=.45\textwidth]{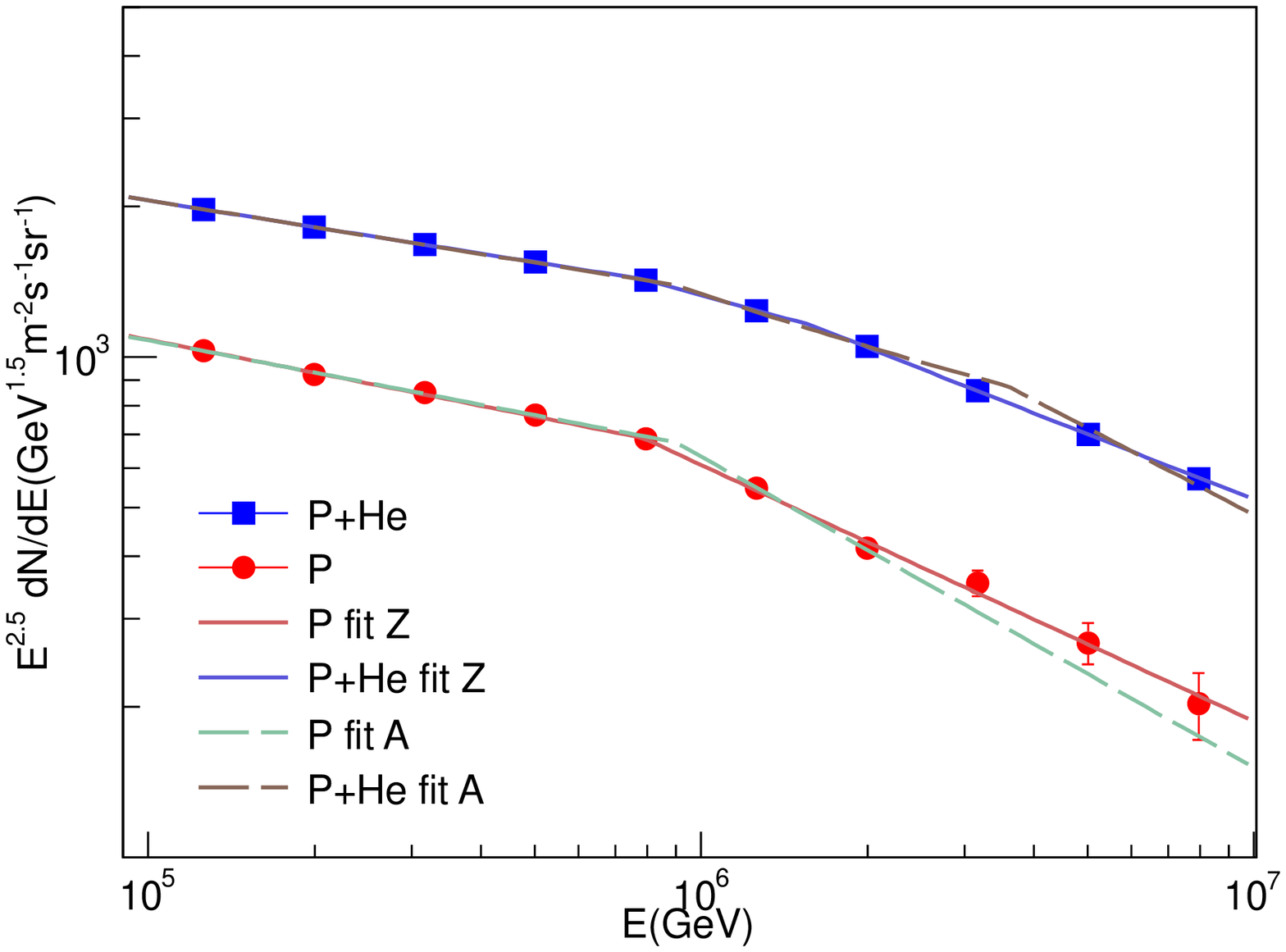}
\includegraphics[width=.45\textwidth]{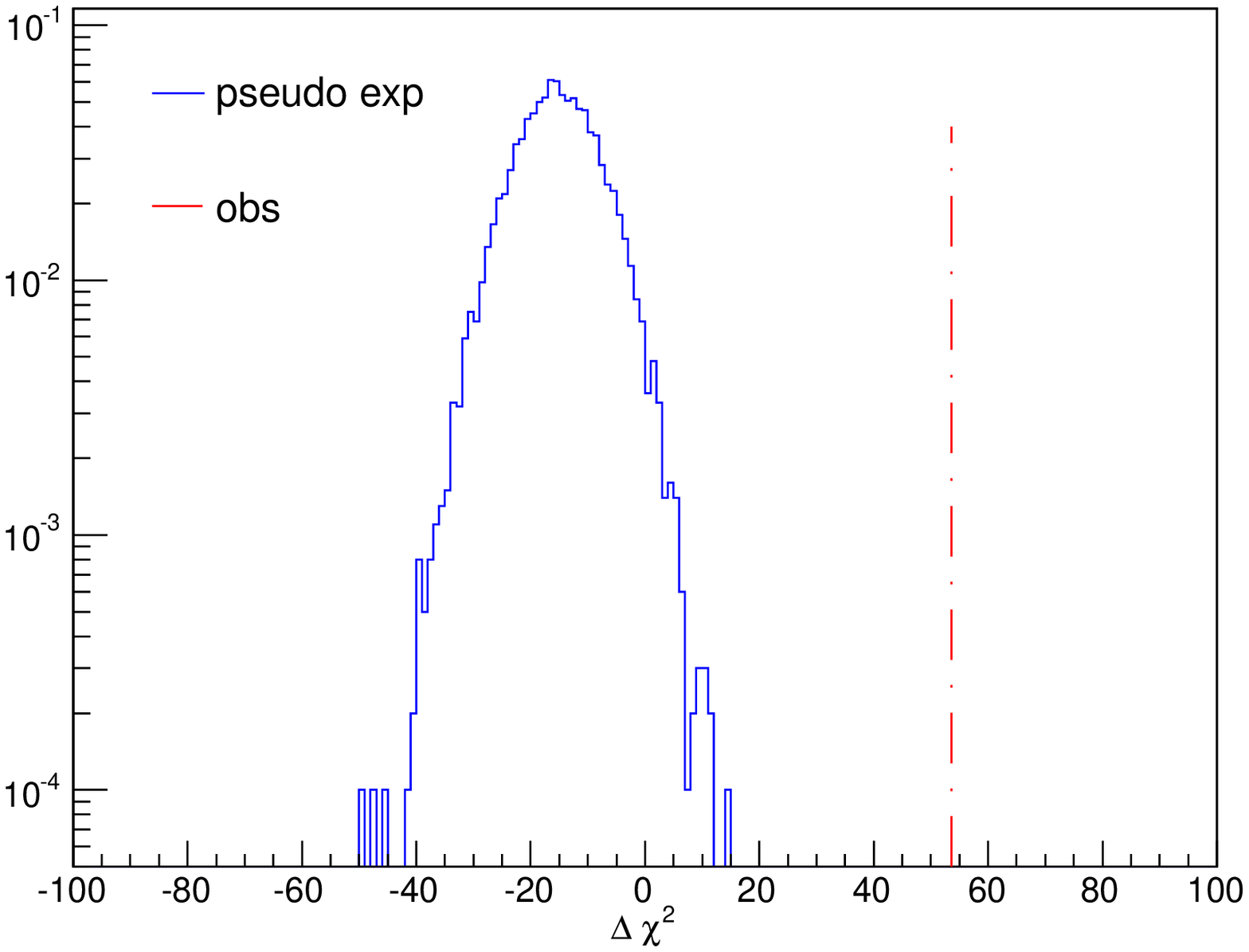}
\caption{left: One sampled pseudo experiment and the corresponding fitting results under the Z-dependent fitting of the X model. The blue dots are the sampled light component spectrum, and the red dots are the sampled proton spectrum. Solid lines are the Z-dependent model ($H_0$) fitting results, and dashed lines are the A-dependent model ($H_1$) fitting result. right: The distribution of the test-statistics $\triangle \chi^2$ (blue) and the observation $\triangle \chi^2_{obs}$ (red dash-doted line) from one sampled observation}
\label{fig:X_model_H0}
\end{figure*}
\end{center}

The results from one of the pseudo experiment under the Z-dependent fitting are shown in the left panel Fig. \ref{fig:X_model_H0}. Different from the observation, the pseudo experiment generates the knee of the He at about 2 PeV and the A-dependent fitting at the knee of He exceeds this energy. The collect of the $\triangle \chi^2$ from all the pseudo experiments and the observation $\triangle \chi^2_{obs}$ are shown in the right panel of Fig. \ref{fig:X_model_H0}. And $\triangle \chi^2$'s distribution shows a well-defined Gaussian shape.  The test-statistcs of the observation separates far from the pseudo experiments, which means the actual model is easy to be recognized. The significance is calculated by dividing the bias $(\triangle \chi^2_{obs} - <\triangle \chi^2>)$ by the root-mean square (RMS) of the distribution of the $\triangle \chi^2$. Averaging the significance over the total 500 sampled observations, the mean significance under the X model is about 10.7 $\sigma$.
\par
Although a prominent significance is obtained under the X model, an essential part is attributed to the intrinsic divergence from the empirical function in Eq. \ref{power_law} of the fine structures in the X model. As shown in Fig. \ref{fig:X_model_obs}, the bump features appear at the break points of both the P and Helium knees, which are not afford to described by the empirical function. In order to eliminate these divergences, we generalize the prediction by parameterizing the X model with the double power-law function in Eq. \ref{power_law}, and denote this new prediction as the GX model. The fitting parameters are listed in Tab. \ref{tab:GX_model}.
\begin{table}
\centering
\caption{\label{tab:GX_model} The fitting parameters of the GX model.}
\begin{tabular}{ccccccc}
  \toprule
  \hline
  parameters                & P & He \\
  \hline
  $\Phi_0$ ( $m^{-2} \cdot s^{-1} \cdot sr^{-1} \cdot TeV^{-1}$ )              & 0.086 & 0.058 \\
  $E_0$ (TeV) & 807 & 4$\times$807 \\
  $\gamma_1$    & -2.7 & -2.64 \\
  $\gamma_2$    & -3.04 & -3.33 \\
  \hline
  \bottomrule
\end{tabular}
\end{table}
\par
\begin{center}
\begin{figure*}
\centering
\includegraphics[width=.45\textwidth]{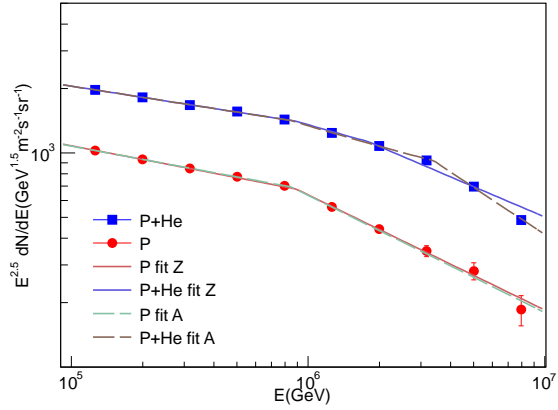}
\caption{One pseudo experiment and the corresponding fitting results under the GX model.}
\label{fig:GX_model_obs}
\end{figure*}
\end{center}

Implementing the same analysis procedure under the GX model, the averaged exclusion significance of the imaginary model is obtained about 6.6 $\sigma$, which is less than the X model by 4.1 $\sigma$. The difference is accordingly caused by the elimination from the fine structures. One of the sampled observation is plotted in the Fig. \ref{fig:GX_model_obs}. It should be pointed that the GX model only describes the A-dependent knee structure, and is independent from different dynamics implied among such kind of interpretations. Thus the result derived from the GX model is suitable to represent all the A-dependent models with He-dominant knee.
%As mentioned above, the A-dependent model is often associated with the new interaction channel appear at the knee scale. Our previous work proposed a new light particle abundant in the Galaxy, together with its threshold interaction with CRS, to explain the knee structure. Hence, it is appropriate to adopt this model as representative of such model in the latter analysis.
%\par

\subsection{Z-Dependent Model Analysis}
From the astrophysical point of view, the origin of the knee, whether induced by the acceleration limit or the leakage from the Galaxy, is compactly associated with the ambient magnetic field. In those mechanisms, the charged part of the nuclei is the participant and leaves the neutral part as the spectator. The Z-dependent knee is commonly adopted and supported by some experiments, such as KASCADE \cite{2012MSAIS..19...49K}, GAMMA \cite{2007APh....28..169G} and so on. On the other hand, it is also found that the spectral knee is a sharp structure \cite{2008ApJ...678.1165A}, which is not favoured by the smooth-knee prediction from the traditional diffusion model \cite{2009arXiv0906.3949E}. In order to account the sharp knee structure under the Z-dependent prediction, we adopt the fitting result from the GAMMA experiment \cite{2014PhRvD..89l3003T} as a representative, which extracts the knee of the P and He at 3.2 and 6.4 PeV respectively.
\par
\begin{center}
\begin{figure*}
\centering
\includegraphics[width=.45\textwidth]{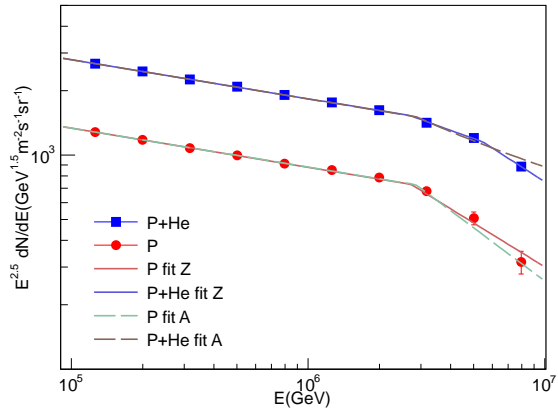}
\caption{One pseudo experiment and the corresponding fitting results under the GAMMA model. The blue dots are the sampled light component spectrum, and the red dots are the sampled proton spectrum. Solid lines are the Z-dependent model ($H_1$) fitting results, and dashed lines are the A-dependent model ($H_0$) fitting result.}
\label{fig:GAMMA_model_obs}
\end{figure*}
\end{center}

The double power-law function in Eq. \ref{power_law} is also used to depict the model hypotheses. One observation and relevant fittings are shown in the Fig \ref{fig:GAMMA_model_obs}. It should be emphasized that the prediction at the knee from the GAMMA model is proton-dominant, while the analysis about the former A-dependent knee based on the He-dominant assumption. This uncertainty about the dominant component causes a significant variance of the testing results. In this scenario, the knee of the He under the $H_0$ fitting is beyond the concerned energy range, which is expected to be $\sim$ 12 PeV.
\par
We find that the corresponding distribution of $\triangle \chi^2$ is not the Gaussian shape and most fitting results from the two model are same while only 5\% of the samples show a slight difference. As mentioned above, the pseudo experiments can exhibit only one spectral break for the light components, which is contributed by the proton. The statistical information can be extracted by integrating the portion of $\triangle \chi^2$ larger than the observation $\triangle \chi^2_{obs}$, which is defined as
\begin{equation}
p-value = P(\triangle \chi^2 \geq \triangle \chi^2_{obs}|H_0)
\end{equation}
And we derive that the averaged p-value is about 0.023, corresponding to 2 $\sigma$.
\par
Besides, another model described by Horandel \cite{2003APh....19..193H} also depicts the Z-dependent knee but a smooth structure. This model is adopted in the analysis as well. Notice that the double power-law function is hard to describe a smooth knee, the poly-gonato model \cite{2003APh....19..193H} with an extra sharpness parameter $\epsilon_c$ is used instead, which is formed as
\begin{equation}
\Phi(E) = \Phi_0 \left( \frac{E}{1 \ TeV} \right)^{\gamma_1} \left[ 1 + \left( \frac{E}{E_{cut}} \right)^{\epsilon_c} \right]^{(\gamma_2 - \gamma_1)/\epsilon_c}
\label{poly_gonato}
\end{equation}
\par
In the Eq. \ref{poly_gonato}, the spectral index is $\gamma_1$ below the knee energy $E_{cut}$ and $\gamma_2$ above, where the intermediate region around the knee depends on the sharpness parameter $\epsilon_c$, which determines the spectral transition rate. Similar to the results of the GAMMA model, the predictions between the two models are found barely no difference. One common feature of them is the P-dominant assumption around 3$ \sim$ 5 PeV, so the fitted knee of the He under the $H_0$ hypothesis is out of energy range as well. As a result, poor exclusion capability is obtained only 1 $\sigma$. In general, due to the limitation from the concerned energy range, our analysis losses the significance for classifying the P-dominant model, whether it is Z-dependent or A-dependent. The measurement extending to higher energies is thus essential in discriminating the knee models.

\section{Conclusion And Discussion}
Realizing that different interpretations about the origin of the knee correspond to different spectral shapes, where the acceleration or the propagation origin result in the Z-dependent knee, and many of the interaction models with new physics results in the A-dependent knee, precise measurement for the individual component is important. Benefit from the merit of the high altitude and the hybrid detection method, the forthcoming LHAASO experiment is expected sensitive to the individual components. In this work, we investigate the capability of LHAASO in distinguishing these knee models. In the consideration of the energy range $10^5 \ \sim \ 10^7$ GeV with 3-year observation statistics, we find the the Z-dependent hypothesis can be excluded at the significance of 6.6 $\sigma$ under the A-dependent knee (He-dominant), while the A-dependent hypothesis mixes with the Z-dependent knee and is harder to be excluded with the significance only 2 $\sigma$.
\par
The influence of the systematic uncertainty can be addressed if we attribute the major part to the energy calibration and the detecting efficiency. These concerned systematic uncertainties lead to the integral shift of the spectrum along the axes. Due to the only acting factor in this analysis is the the ratio of the knee energy of the He and P, which is split into 4 and 2 with respect to different knee models, the integral shift of the spectrum is not expected to influence the ratio significantly. Thus, this test is insensitive to the variance of the systematic uncertainties, which is a unique advantage.
\par
The lack of significant in recognition of the Z-dependent knee in this analysis is mainly due to the relative narrow energy band around the knee and the P-dominant assumption. Other modes of LHAASO that focus the higher energies is required to further determine the spectral index above the He's knee. Meanwhile latest measurement from ARGO \cite{2014arXiv1408.6739D, 2016NPPP..279....7M}, ARGO+WFCT \cite{2015PhRvD..92i2005B} and $AS\gamma$ \cite{2006PhLB..632...58T} also find some hints that the knee of the light components occurs at hundreds TeV, which corresponds to higher flux at the knees of the P and He. If this observation is confirmed by further LHAASO experiment, the analysis will fall into the concerned energy range naturally and can be performed much more conveniently with higher statistics.

\section{Acknowledgements}
This work is supported by the National Key R\&D Program of China (number 2018YFA0404200), the Natural Sciences Foundation of China (numbers 11575203, 11635011).

\end{document}